\documentstyle[11pt,epsf]{article}
\font\elevenbf=cmbx10 scaled\magstep 1
\font\elevenrm=cmr10 scaled\magstep 1
 1

\renewenvironment{thebibliography}[1]
 { \elevenrm
   \begin{list}{\arabic{enumi}.}
    {\usecounter{enumi} \setlength{\parsep}{0pt}
     \setlength{\itemsep}{3pt} \settowidth{\labelwidth}{#1.}
     \sloppy
    }}{\end{list}}
\begin{document}
\noindent
\thispagestyle{empty}
\renewcommand{\thefootnote}{\fnsymbol{footnote}}
\begin{flushright}
{\bf TTP95-17}\footnote{The complete postscript file of this
preprint, including figures, is available via anonymous ftp at
ttpux2.physik.uni-karlsruhe.de (129.13.102.139) as
/ttp95-17/ttp95-17.ps or via www at
http://ttpux2.physik.uni-karlsruhe.de/cgi-bin/preprints/
Report-no: TTP95-17.}\\
{\bf June 1995}\\
hep-ph/9507255\\
\end{flushright}
\begin{center}
\begin{bf}
 \begin{large}
 HADRON RADIATION IN \end{large}
 \begin{Large}$\tau$\end{Large}
\end{bf}
 \begin{large}
 PRODUCTION AND\\
 THE LEPTONIC $Z$ BOSON DECAY RATE\footnote{Work
 supported by BMFT 056 KA 93P.}\\
 \end{large}
  \vspace{0.5cm}
  \begin{large}
   A.H. Hoang\footnotemark[4], J.H. K\"uhn\footnote{Present address:
SLAC, Stanford University, Stanford, CA 94309.} and
   T. Teubner\footnote{e--mails: hoang@ttpux2.physik.uni-karlsruhe.de
 and tt@ttpux2.physik.uni-karlsruhe.de}
 \\
  \end{large}
  \vspace{0.1cm}
Institut f\"ur Theoretische Teilchenphysik, Universit\"at Karlsruhe,\\
D--76128 Karlsruhe, Germany \\
  \vspace{0.7cm}
  {\bf Abstract}\\
\vspace{0.3cm}
\noindent
\begin{minipage}{15.0cm}
\begin{small}
Secondary radiation of hadrons from a tau pair produced
in electron positron collisions may constitute an
important obstacle for precision measurements of
the production cross section and of branching ratios.
The rate for real and virtual radiation is calculated
and various distributions are presented.
For Z decays a comprehensive analysis is performed which
incorporates real and virtual radiation of
leptons. The corresponding results are also given for
primary electron and muon pairs.
Compact analytical formulae are presented for
entirely leptonic configurations.
Measurements of $Z$ partial decay rates which eliminate
all hadron and lepton radiation are about 0.3\% to 0.4\%
lower than totally inclusive measurements, a consequence of
the ${\cal O}(\alpha^2)$ negative virtual corrections which
are enhanced by the third power of a large logarithm.
\end{small}
\end{minipage}
\end{center}
\setcounter{footnote}{0}
\renewcommand{\thefootnote}{\arabic{footnote}}
\vspace{1.2cm}
%
%
%
\noindent
{\bf 1.~Introduction}\\
\par
\noindent
Since the turn on of LEP several years ago experiments have
collected an impressive amount of data, with a combined
sample of more than $10^7$ hadronic and $10^6$ leptonic
events. This enormous event rate has allowed us to measure the
ratio of hadronic versus leptonic events or the leptonic
decay rate with an accuracy of better than two per mill.
Clearly, at this level of precision subtle aspects of
radiative corrections have to be taken into account. One specific
example is the $Z$ decay rate into a lepton pair: Details of the
treatment of additional radiation of a pair of soft leptons or of
a soft hadronic system have a strong impact on the predicted rate.
Formally the rate of these events is of
order $(\alpha/\pi)^2\approx 5\cdot 10^{-6}$.
In reality, however, these events are enhanced by the third
power of a large logarithm, which essentially compensates
one factor of $\alpha/\pi$. This positive contribution is to
a large extent cancelled by the corresponding virtual
correction, leaving a small positive shift in the inclusive rate.
The dominant part of this shift can be incorporated into the
corrections of order $\alpha/\pi$ by employing the running QED
coupling $\alpha(Q^2)$, the remainder is truly negligible.
However, in an experimental analysis which eliminates events with
secondary radiation the large negative virtual corrections must
be taken into account.
\par
With this motivation in mind
we present in this paper a comprehensive study of virtual and
real leptonic and hadronic radiation. Particular emphasis is
put on the radiation of hadrons from a primary $\tau$ lepton pair.
This is the most difficult reaction to evaluate, since three
independent scales $s$,  $m_\tau$ and $m_{had}$ enter the
calculation. The results for the radiation of hadrons from
an electron or muon pair are significantly simpler:
The lepton mass is far smaller than the threshold for hadron
production and can be safely set to zero throughout the calculation.
Results for this latter case have been given in~\cite{KKKS} for
virtual corrections in the high energy approximation.
The corresponding case for real final state radiation with a
massless primary lepton has been solved in analytic form
in~\cite{HJKT1,HJKT2}. The virtual and real radiation of a lepton
pair off a massive primary lepton has been evaluated for
arbitrary $s$ in~\cite{HKT1} for the special case that the mass of the
secondary lepton is far smaller than all other scales of the problem.
In this paper we complete this topic by considering
radiation of hadrons from a primary $\tau$ pair. Since large
data samples of $\tau$ pairs are also collected at lower
energies, e.g. at around 10 GeV, this case also is
scrutinized. However, it is observed that radiation is far
less important in this low energy region, an obvious
consequence of the less prominent role of the ``large'' logarithms.
Hadronic radiation off primary $\tau$ pairs also plays a
special role from a purely practical point of view: $\tau$
leptons decay into hadrons (plus a neutrino) and hence
events with additional hadron radiation might well be
considered experimentally as hadronic final states, since
they have larger hadronic multiplicity and the invariant
mass of the hadronic subsystems exceeds $m_\tau$. By the
same token, the measurement of $\tau$ decays into large
multiplicity final states could in principle be influenced
by inappropriate assignment of the radiative events. Last but
not least, signals of ``new'' physics could be faked by a
misinterpretation of these Standard Model reactions.
\par
The final state radiation discussed  in this paper is
complementary to the initial state radiation treated in~\cite{KKKS,JSM}.
No interference terms are present between radiation from the initial
and final state as long as the (dominant) QED reactions are considered
and axial vector couplings from neutral current reactions are ignored.
Final state radiation of lepton (and quark) pairs has also
been treated in~\cite{vdB} in the context
of a Monte  Carlo Program. For hadron radiation this
approach turns out to be inadequate, since the reaction is
dominated by low mass hadronic systems. For lepton radiation our
results are lower than those of~\cite{vdB} by about 15\%.
An important ingredient for the evaluation of real and virtual
hadronic radiation is a proper description of $R(s)$, which has to be
taken from experiment. In our approach we employ the most recent
parametrization of data taken from~\cite{jegerl}.
\par
The plan of this paper is as follows: In section 2 the
``master formula'' for calculation of the total rate is
presented, which is at the same time a convenient starting
point for a variety of differential distributions. The case
of real radiation with primary $\tau$-leptons and secondary
hadrons is discussed in some detail, considering high
energies ($\sqrt{s}=M_Z$) as well as $\sqrt{s}=10$~GeV
(characteristic for a $B$ factory) and $4.2$~GeV (relevant
for a $\tau$-charm factory).
Predictions for the total
rate are presented, as well as distributions which are differential
with respect to the mass of the hadronic system and its
energy. The corresponding virtual radiation is calculated
in section 3 for the same c.m. energies and at the $\tau$ pair
threshold. Corrections are given for
the Dirac and Pauli form factor, as well as for the total rate.
Section 4 is concerned with a detailed discussion of $Z$ decays.
Results for electron, muon and tau as primary
leptons are presented both for hadronic and leptonic
radiation. The corresponding predictions for virtual
radiation are also collected. It is demonstrated that
these virtual corrections reduce the exclusive leptonic rate
of the $Z$ boson by several per mill.
Section 5 contains a summary and \pagebreak[4]
our conclusions.
\par
\vspace{1cm}\noindent
{\bf 2.~Real radiation}\\
\par
\noindent
The calculation of the ratio $\sigma_{\tau^+ \tau^- had}/
 \sigma_{pt}$ (where $\sigma_{pt} = 4\,\pi\,\alpha^2/3\,s$) can be
reduced to the numerical evaluation of the two dimensional integral
\begin{eqnarray}
\lefteqn{
 R_{\tau^+ \tau^- had} \equiv
 \frac{\sigma_{\tau^+ \tau^- had}}
      {\sigma_{pt}} =
 \frac{1}{3} \,
 \left(\frac{\alpha}{\pi}\right)^2\,
 \int\limits_{4m_{\pi}^2}^{(\sqrt{s}-2m_{\tau})^2} \frac{{\rm d}s'}{s'}
 \, R_{had}(s') \, F(s'/s) \,,} \label{masterreal}
 \\[1mm]
\lefteqn{
F(z) = 4
 \int\limits_{4m_{\tau}^2/s}^{(1-\sqrt{z})^2} {\rm d}y \,
 \left\{
 - \,
 \sqrt{1-\frac{4m_{\tau}^2}{s\,y}} \, \Lambda^{1/2}(1,y,z)\,\left[ \,
  \frac{1}{4} + \frac{\frac{2m_{\tau}^2}{s}+\frac{4m_{\tau}^4}{s^2}+
                      \left(1+\frac{2m_{\tau}^2}{s}\right)z}
                     {(1-y+z)^2 -
                      \left(1-\frac{4m_{\tau}^2}{s\,y}\right) \,
                     \Lambda(1,y,z)}
  \, \right]\,\right.}\nonumber\\[1mm]  & & \left. +
 \frac{\frac{2m_{\tau}^4}{s^2}+\frac{m_{\tau}^2}{s}(1-y+z)-\frac{1}{4}
           (1-y+z)^2-\frac{1}{2}(1+z)y}{1-y+z}\,
  \ln\frac{1-y+z-\sqrt{1-\frac{4m_{\tau}^2}{s\,y}}\,\Lambda^{1/2}(1,y,z)}
          {1-y+z+\sqrt{1-\frac{4m_{\tau}^2}{s\,y}}\,\Lambda^{1/2}(1,y,z)}
 \mbox{} \,  \right\} \nonumber
\end{eqnarray}
where
$$
\Lambda(1,y,z)=1+y^2+z^2-2(y+z+y\,z) \,.
$$
$F(s'/s)$ can be interpreted as the normalized cross section
for $\tau$ pair production with the additional emission of a
vector boson of mass $\sqrt{s'}$. The integral in (\ref{masterreal})
sums up all hadronic contributions of mass $\sqrt{s'}$ weighted
by $R_{had}(s')$.
\begin{figure}
\begin{center}
\leavevmode
\epsfxsize=6.5cm
\epsffile[200 360 360 480]{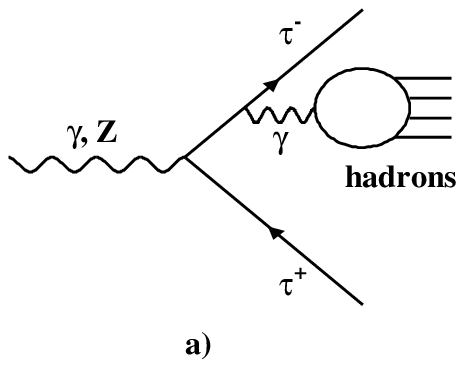}
\leavevmode
\epsfxsize=6.5cm
\epsffile[200 360 360 480]{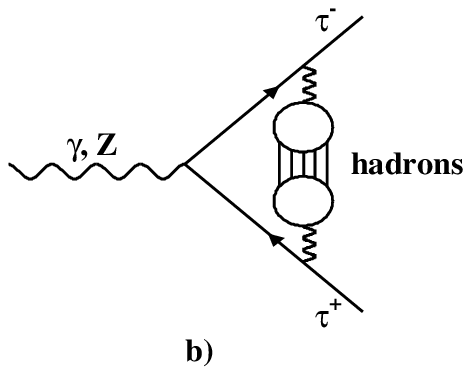}
\vskip -3mm
\caption[]{{\em Typical diagrams for real and virtual emission of
hadrons in $\tau$ pair production.}}
\label{figgraphs}
\end{center}
\end{figure}
The numerical
treatment of this integral allows us to use experimental input for the
function $R_{had}(s')$ and avoids the poor approximations of a parton
model inspired calculation.
In contrast to \cite{HKT1} --- there we solved
the integrals analytically in the case where the ``massive'' photon
splits into a pair of fermions with mass far smaller
than $m_{\tau}$ --- $F(s'/s)$ is calculated numerically.
We would like to emphasize that (\ref{masterreal}) contains no restriction
on the mass values and is thus also applicable to completely different
kinematical situations.
In the limit of vanishing external mass $m_{\tau}=0$ we
recover eq.~(2) of \cite{HJKT1} for $F(s'/s)$.
\par
The physical situation described by (\ref{masterreal}) consists typically
of a $\tau$ pair with large invariant mass and a hadronic state with
low invariant mass (see also the discussion of our results below).
The hadrons are emitted relatively
collinear with one of the $\tau$ leptons.
The reverse configuration, where a $\tau$ pair is radiated off the
corresponding quark via a virtual photon could be treated with a
similar technique. This final state would consist mainly of quite
soft and collinear virtual photons leading to a $\tau$ pair of low
invariant mass and is therefore easily distinguished from the
situation we focus on in this letter.
As far as the total rate is concerned,
the interference term between these two amplitudes vanishes
if the production is induced
by the (electromagnetic or neutral) vector current. For axial
vector induced contributions the interference term is different
from zero. However, this term does not exhibit a mass singularity
in the limit of vanishing quark or leptonic mass and
thus is not enhanced by a large logarithm. It can be
neglected in all cases of interest. In the limit of vanishing
fermion mass this interference term has been calculated in \cite{KK}.
\par
For most of this paper we will discuss the total rate for
secondary radiation. However, eq.(\ref{masterreal}) contains
implicitly also information on single and double differential
distributions. The integration variable $s^\prime$ denotes
the squared hadronic mass. The square of the invariant mass
of the primary $\tau^+\tau^-$ system, on the other hand, is given by
$y\cdot s$, which can be easily related to the energy
$E_{\scriptsize had}$ of the hadronic system.
The single differential distribution
${\rm d} R_{\tau^+ \tau^- had}/{\rm d} s^\prime$
is trivially obtained from~eq.(\ref{masterreal}). The energy
distribution of the radiated hadronic system is given by
\begin{eqnarray}
\lefteqn{
\frac{{\rm d} R_{\tau^+ \tau^- had}}{{\rm d} E_{\rm\scriptsize had}}  = }
  \nonumber\\[1mm]
& & \frac{4}{3}\, \left(\frac{\alpha}{\pi}\right)^2\,
 \int\limits_{s'_{\scriptsize min}}^{E_{\scriptsize had}^2}
   \frac{{\rm d}s'}{s\,s'}
 \, R_{had}(s') \,
 \left\{ - \,
 \sqrt{1-\frac{4m_{\tau}^2}{s\,y}} \,
 \sqrt{E_{\scriptsize had}^2-s^\prime} \,\left[ \,
  1 + \frac{2m_{\tau}^2+\frac{4m_{\tau}^4}{s}+
               \left(1+\frac{2m_{\tau}^2}{s}\right)s^\prime}
           {E_{\scriptsize had}^2-
            \left(1-\frac{4m_{\tau}^2}{s\,y}\right) \,
            (E_{\scriptsize had}^2-s^\prime)}
  \, \right]\,\right.\nonumber\\[1mm] & & \left.\quad +\,
 \frac{s}{E_{\scriptsize had}}\left(
   \frac{2m_{\tau}^4}{s^2}+
   \frac{2 E_{\scriptsize had} m_{\tau}^2}{s^{3/2}}-
   \frac{E_{\scriptsize had}^2}{s}-\frac{1}{2}(1+\frac{s^\prime}{s})y
 \right)\,
  \ln\frac{E_{\scriptsize had}-\sqrt{1-\frac{4m_{\tau}^2}{s\,y}}\,
              \sqrt{E_{\scriptsize had}^2-s^\prime}}
          {E_{\scriptsize had}+\sqrt{1-\frac{4m_{\tau}^2}{s\,y}}\,
              \sqrt{E_{\scriptsize had}^2-s^\prime}}
 \mbox{} \,  \right\}
\label{diffenergyreal}
\end{eqnarray}
with
\begin{equation}
s\,y = s+s^\prime-2 E_{\scriptsize had} \sqrt{s} \qquad
\mbox{and}\qquad
s'_{\scriptsize min} = {\rm max}\left(4 m_\tau^2-s+2 E_{\scriptsize
had}\,\sqrt{s}\,,\, 4 m_\pi^2\right)\,.
\end{equation}
These formulae will be useful below.
\par
In the numerical evaluation of (\ref{masterreal}) we use experimental data
for $R_{had}(s')$ as presented in \cite{jegerl}\footnote{We thank
F.~Jegerlehner and S.~Eidelman for providing their
compilation of data and discussing its proper treatment.}.
A convenient approach adapted to the experimental procedure is to decompose
the hadronic final states into continuum and resonance contributions.
Following \cite{jegerl}, the two pion contribution ($\widehat{=} \rho$) is
taken into account together with the continuum by a direct integration of
data points. The most important narrow resonances are added separately
using the narrow width approximation\footnote{Masses and electronic widths
are taken from \cite{PDG1994}.}, which is accurate enough for our
purposes. More details about $R_{had}(s')$ are given in the appendix.
\par
\begin{figure}
\begin{center}
\leavevmode
\epsfxsize=12cm
\epsffile[120 300 450 530]{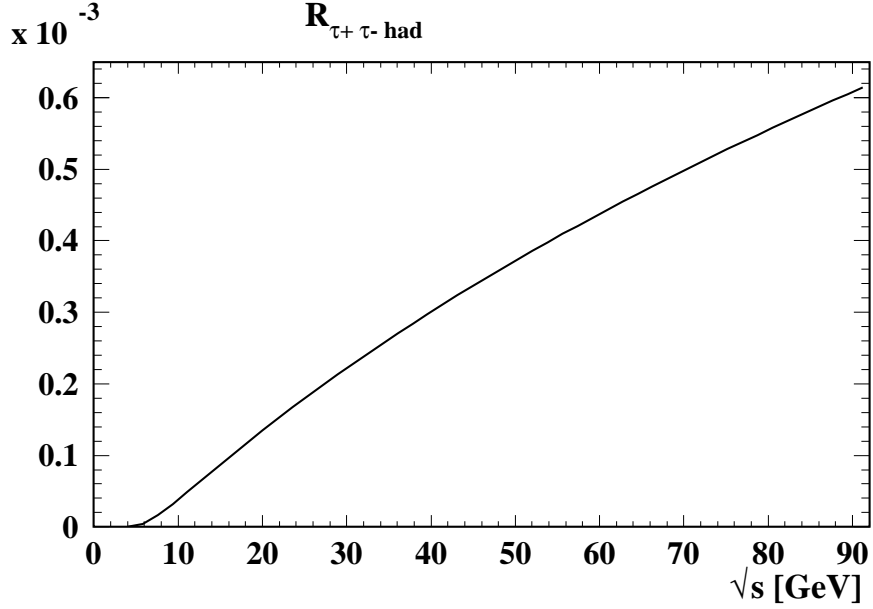}
\vskip -3mm
\caption[]{{\em Normalized cross section
$R_{\tau^+ \tau^- had}$.}}
\label{scanr}
\end{center}
\end{figure}

In Table~\ref{tab1}
we display the continuum plus $\rho$ and the resonance
contributions separately for different center of mass energies
$\sqrt{s} \le M_Z$.
The contribution from a virtual $Z$ as intermediate state remains
unimportant up to $\sqrt{s} \approx M_Z$. They are strongly
suppressed for obvious kinematical reasons as can be easily seen
in Table~\ref{tab1} (see also discussion below). This in turn
justifies neglecting radiation through the virtual $Z$.
In Fig.~\ref{scanr}\ $\ R_{\tau^+ \tau^- had}$
is plotted for the range $2 m_{\tau} < \sqrt{s} < M_Z$.
The rise of $R_{\tau^+ \tau^- had}$ and
$R_{\tau^+ \tau^- l^+ l^-}$ for increasing $\sqrt{s}$ is easily understood
from the leading term in the high energy limit proportional to the
third power of the logarithm of $s$, i.e.
$\ln^2(m_{had}^2/s)\ln(m_\tau^2/s)$, $m_{had}$ being
a characteristic hadronic mass scale of ${\cal O}(300\ {\rm MeV})$. (For
leptonic radiation these logarithms can be calculated explicitly,
see eq.~(\ref{mismallreal}).)
\par
\begin{table}
\begin{center}
\begin{tabular}{|c||c|c|c|} \hline
$\sqrt{s}$ & $4.2$ GeV       & $10$ GeV        & $M_Z$           \\ \hline
\hline
$\sqrt{s'_{max}}$
           & $0.646$ GeV     & $6.446$ GeV     & $87.63$ GeV     \\ \hline
\hline \hline
Channel    & \multicolumn{3}{|l|}{}                              \\ \hline
$2 \pi$    & $5.11\cdot 10^{-9}$ & $2.07\cdot 10^{-5}$ & $1.81\cdot 10^{-4}$
\\ \hline
$\omega$   & --              & $2.19\cdot 10^{-6}$ & $2.08\cdot 10^{-5}$ \\
\hline
$\phi$     & --              & $2.81\cdot 10^{-6}$ & $3.22\cdot 10^{-5}$ \\
\hline
$J/\psi$   & --              & $3.54\cdot 10^{-7}$ & $2.82\cdot 10^{-5}$ \\
\hline
$\Upsilon$ & --              & --              & $8.17\cdot 10^{-7}$ \\
\hline
\multicolumn{4}{|l|}{Continuum (with $R_{had}(s')$ from experiment):} \\
\hline
$0.81$ up to $4.98$ GeV
           & --              & $1.22\cdot 10^{-5}$ & $2.59\cdot 10^{-4}$ \\
$4.98$ up to $10$ GeV
           & --              & $7.13\cdot 10^{-9}$ & $0.60\cdot 10^{-4}$ \\
$10$ up to $25$ GeV
           & --              & --              & $0.30\cdot 10^{-4}$\\
$25$ up to $40$ GeV
           & --              & --              & $0.03\cdot 10^{-4}$\\
\hline
\multicolumn{4}{|l|}{Continuum (with perturbative $R_{had}(s')$)} \\
\hline
$\sqrt{s'} > 40$ GeV
           & --              & --              & $3.70\cdot 10^{-7}$ \\
\hline \hline
$R_{\tau^+ \tau^- had}$
           & $5.11\cdot 10^{-9}$ & $3.82\cdot 10^{-5}$ & $6.14\cdot 10^{-4}$
\\ \hline \hline \hline
$R_{\tau^+ \tau^- e^+ e^-}$
           & $2.75\cdot 10^{-5}$ & $4.05\cdot 10^{-4}$ & $1.82\cdot 10^{-3}$
\\ \hline
$R_{\tau^+ \tau^- \mu^+ \mu^-}$
           & $1.14\cdot 10^{-7}$ & $2.82\cdot 10^{-5}$ & $2.94\cdot 10^{-4}$
\\ \hline
$R_{\tau^+ \tau^- \tau^+ \tau^-}$
           & --                  & $3.07\cdot 10^{-8}$ & $3.13\cdot 10^{-5}$
\\ \hline \hline \hline
   Sum     & $2.76\cdot 10^{-5}$ & $4.71\cdot 10^{-4}$ & $2.76\cdot 10^{-3}$
\\ \hline
\end{tabular}
\caption[]{\label{tab1}
{\em Contributions to $R_{\tau^+ \tau^- had}$ and values of
$R_{\tau^+ \tau^- l^+ l^-}$ ($\,l = e,\,\mu\,,\tau$)
for center of mass
energies $\sqrt{s} = 4.2$ GeV/$10$ GeV/$M_Z$. $\sqrt{s'_{max}}$ is the upper
limit of the phase space integral (\ref{masterreal}). }}
\end{center}
\end{table}
As an interesting exercise one may apply (\ref{masterreal}) to
$\tau^+ \tau^- q \bar q$ production where the
quarks are again radiated off the tau leptons leaving $\tau$ pairs of
large invariant mass\footnote{In this case one just has to use
$R(s') = N_c \, Q_q^2 \, (1+\frac{2m_q^2}{s'})\,
 \sqrt{1-\frac{4m_q^2}{s'}}$ in the numerical integration.}.
This leads to a parton model like prediction of
$R_{\tau^+ \tau^- had}$. Adopting
the quark mass values of \cite{vdB}, viz.
$m_u=m_d=m_s=300 \ {\rm MeV}$, $m_c=1.5 \ {\rm GeV}$
and $m_b=4.5 \ {\rm GeV}$ and choosing $\sqrt{s} = 10\ {\rm GeV}\,/\,M_Z$
one \pagebreak[4] would predict
\begin{eqnarray}
R_{\tau^+ \tau^- u \bar u}&=&4\,R_{\tau^+ \tau^- d \bar d} =
                             4\,R_{\tau^+ \tau^- s \bar s} =
 1.1\cdot 10^{-5}\,/\,2.1\cdot 10^{-4}\,,
 \nonumber\\
R_{\tau^+ \tau^- c \bar c}&=&
 1.2\cdot 10^{-7}\,/\,5.1\cdot 10^{-5}\,,\nonumber\\
R_{\tau^+ \tau^- b \bar b}&=&
 \qquad 0 \quad\ \  / \,2.6\cdot 10^{-6}\,,\nonumber\\[1mm]
R_{\tau^+ \tau^- had}&=&
 1.7\cdot 10^{-5}\,/\,3.7\cdot 10^{-4}\,.
\label{partonmodel}
\end{eqnarray}
It is obvious that this model underestimates the production rate by
about a factor of two. (The same phenomenon has also been observed
for hadron radiation off muons and electrons \cite{HJKT1}.)
For energies near the $\tau^+ \tau^-$ production
threshold a parton model like prediction is completely impossible
without artificial tuning of the masses of the light quark flavours.
\par
\begin{figure}
\begin{center}
\leavevmode
\epsfxsize=8.6cm
\epsffile[170 320 400 520]{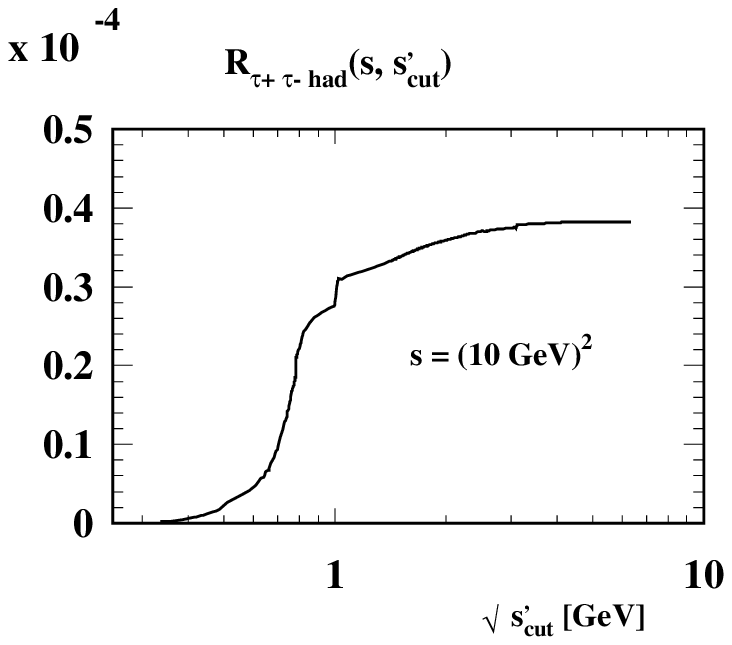}
\leavevmode
\epsfxsize=8.6cm
\epsffile[170 320 400 520]{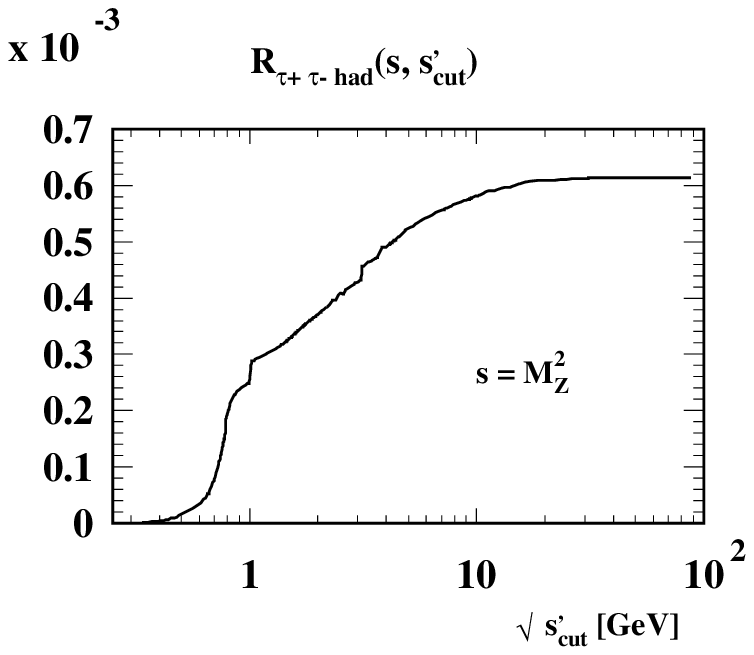}
\vskip -3mm
\caption[]{{\em Dependence of the normalized cross section
$R_{\tau^+ \tau^- had}$ from a cut $s'_{cut}$ for
$\sqrt{s} = 10\ {\rm GeV}$ and $\sqrt{s} = M_Z$.}}
\label{figcut}
\end{center}
\end{figure}
For completeness we also list in Table~\ref{tab1} the rate for the
radiation of lepton ($e,\,\mu,\,\tau$) pairs. The prediction for the
radiation of $e^+ e^-$ and $\mu^+ \mu^-$ is based on the analytical
formula given in \cite{HKT1}, the one for $\tau^+ \tau^- \tau^+ \tau^-$
is obtained through numerical integration. An adequate approximation
(better than 1\%) for
$R_{\tau^+ \tau^- \tau^+ \tau^-}$ at $\sqrt{s} = M_Z$ is presented in
eq.~(\ref{miequalreal}) at the end of section 4.
\begin{figure}
\begin{center}
\leavevmode
\epsfxsize=8.6cm
\epsffile[175 335 400 515]{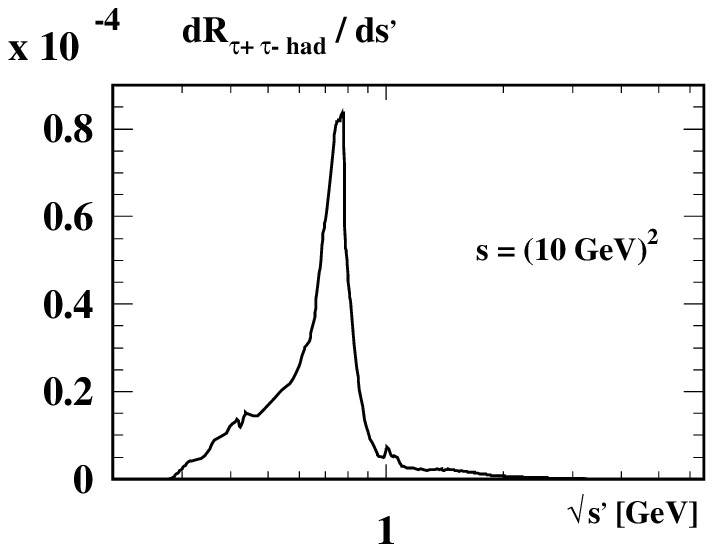}
\leavevmode
\epsfxsize=8.6cm
\epsffile[175 335 400 515]{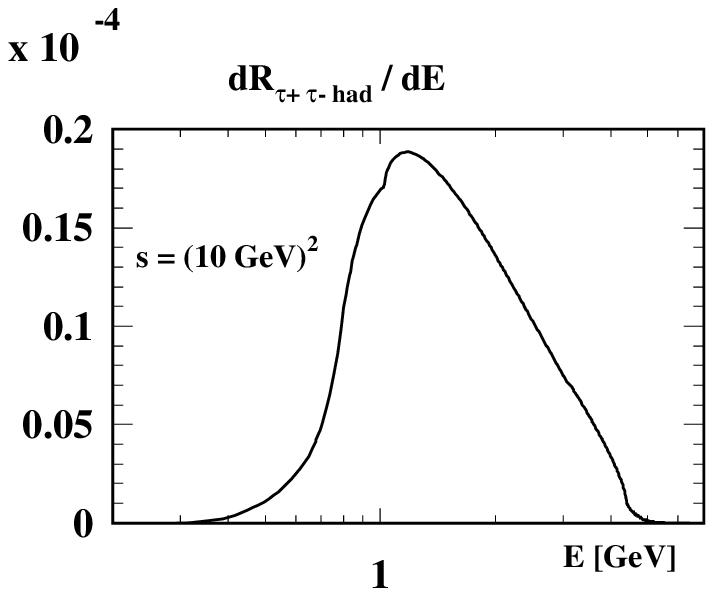}\\
\leavevmode
\epsfxsize=8.6cm
\epsffile[175 335 400 515]{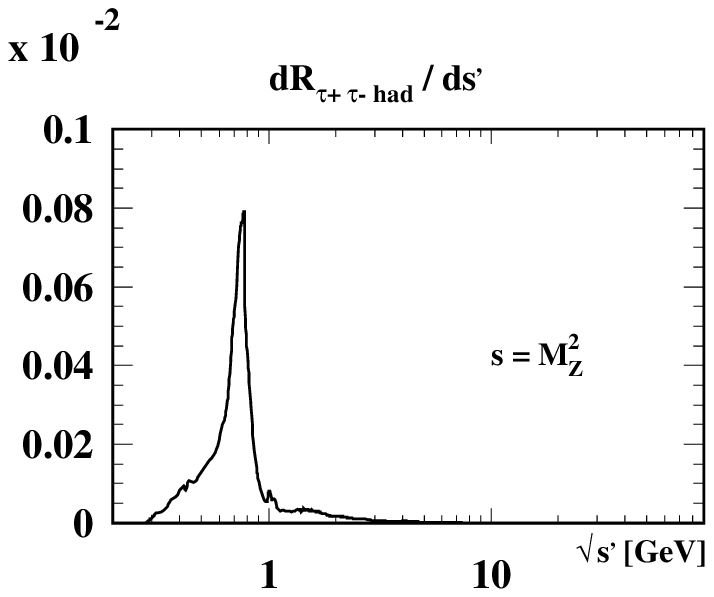}
\leavevmode
\epsfxsize=8.6cm
\epsffile[175 335 400 515]{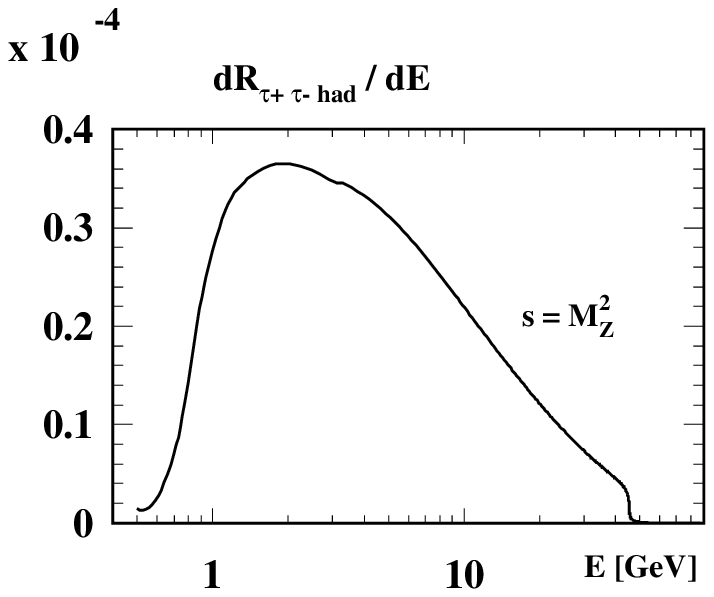}
\vskip -3mm
\caption[]{{\em Differential distributions for real emission of
hadrons in $\tau$ pair production as described in the text.}}
\label{figtauhaddif}
\end{center}
\end{figure}
\par
Table~\ref{tab1}
and Fig.~\ref{scanr} demonstrate that hadronic radiation
may become important for precision measurements of $\tau$ branching ratios
on top of the $Z$. Radiation of $e^+ e^-$ or $\mu^+ \mu^-$ may even be of
relevance in the $10$ GeV region, e.g. at a $B$ meson factory.
\par
\begin{figure}
\begin{center}
\leavevmode
\epsfxsize=8.6cm
\epsffile[175 335 400 515]{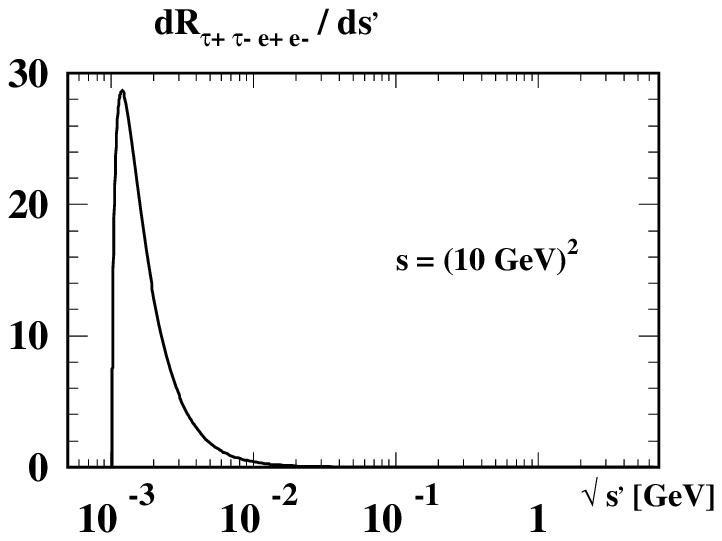}
\leavevmode
\epsfxsize=8.6cm
\epsffile[175 335 400 515]{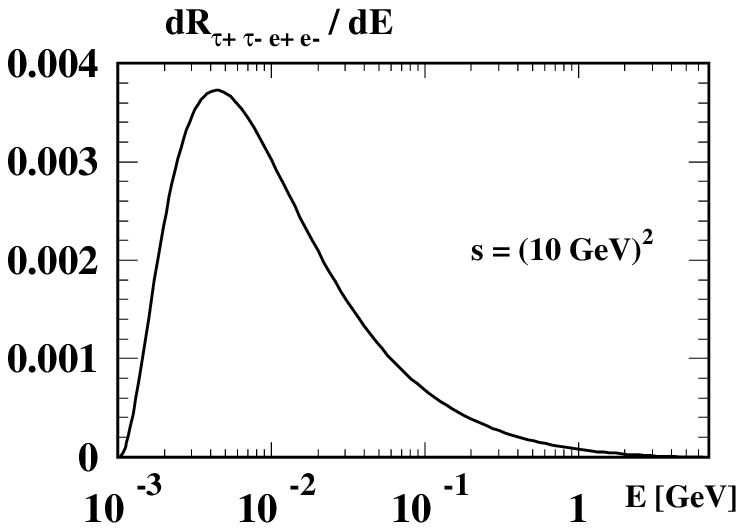}\\
\leavevmode
\epsfxsize=8.6cm
\epsffile[175 335 400 515]{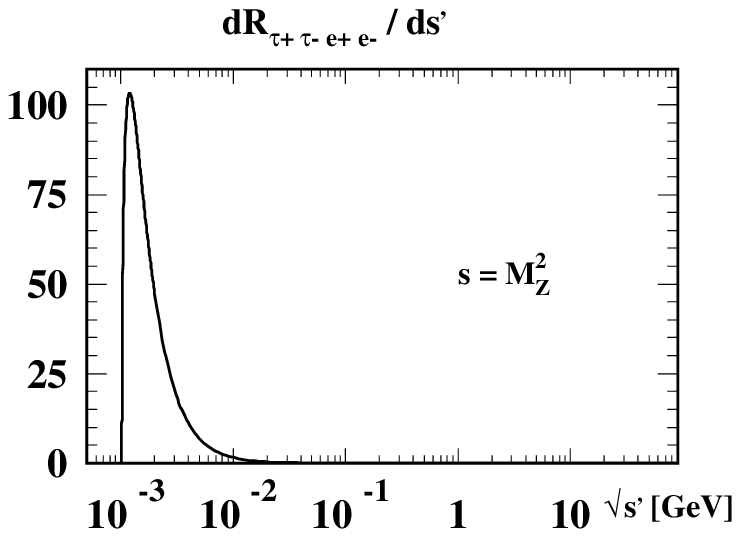}
\leavevmode
\epsfxsize=8.6cm
\epsffile[175 335 400 515]{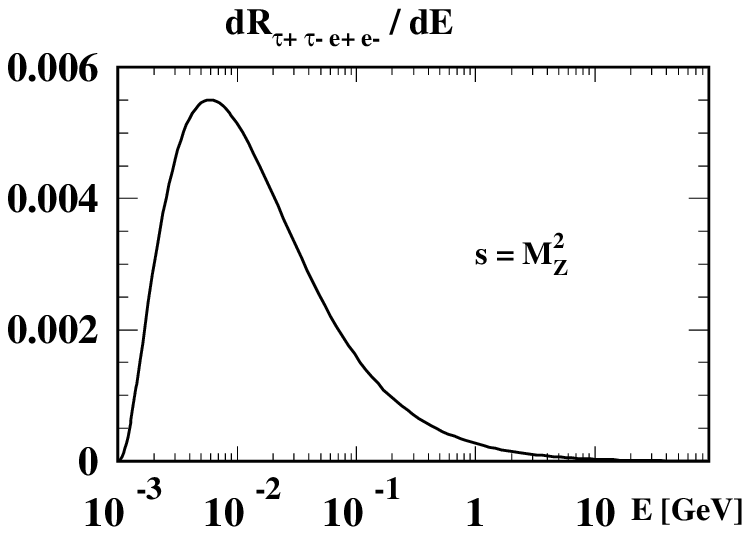}
\vskip -3mm
\caption[]{{\em Differential distributions for real emission of
$e^+ e^-$ in $\tau$ pair production as described in the text.}}
\label{figtauedif}
\end{center}
\end{figure}

It is evident that experimental sensitivity and the probability to
assign $\tau^+ \tau^-$ events with hadronic radiation to exclusive or
inclusive $\tau$ pair production depends strongly on the invariant mass
of the hadronic system. As mentioned above the integral (\ref{masterreal})
is in fact dominated by small $s'$. In the
numerical approach this is easily shown by replacing the upper limit
of the integral in (\ref{masterreal}) by a value
$s'_{cut} < s'_{max} = (\sqrt{s}-2\,m_{\tau})^2$. In
Fig.~\ref{figcut} we display our results for
$R_{\tau^+ \tau^- had}(s,s'_{cut}) =
\sigma_{\tau^+ \tau^- had}(s,s'_{cut})/\sigma_{pt}$  for
$\sqrt{s}=10$ GeV/$M_z$. Even for $Z$ decays 86\% of the integral
originates from $\sqrt{s'} < 5$ GeV and 58\% from
$\sqrt{s'} < m_{\tau}$.
In Fig.~\ref{figtauhaddif} we present the differential distributions
${\rm d} R_{\tau^+ \tau^- had}/{\rm d} s^\prime$ and
${\rm d} R_{\tau^+ \tau^- had}/{\rm d} E_{\scriptsize had}$
for $\sqrt{s} = 10\ {\rm GeV}\,/\,M_Z$
using eqs.(\ref{masterreal}) and (\ref{diffenergyreal}).
Contributions from the narrow resonances are not taken into account
for the distributions ${\rm d} R_{\tau^+ \tau^- had}/{\rm d} s^\prime$ but
can easily be read off from Table~\ref{tab1}.
As expected the distributions are peaked at small $s^\prime$ (already
evident from Fig.~\ref{figcut}) and at fairly small $E_{\scriptsize had}$
illustrating that hadronic radiation is clearly dominated by
the $\rho$--resonance.
It is not inconceivable that these events might
affect precision measurements of multipion decay modes of the~$\tau$.
In view of possible experimental applications we also present in
Fig.~\ref{figtauedif} the differential distributions
${\rm d} R_{\tau^+ \tau^- e^+ e^-}/{\rm d} s^\prime$ and
${\rm d} R_{\tau^+ \tau^- e^+ e^-}/{\rm d} E_{e^+ e^-}$ for the
radiation of electron--positron pairs at $\sqrt{s} = 10\ {\rm GeV}$
and at $M_Z$. These are evidently also peaked at
extremely small values of $s^\prime$ and $E_{e^+ e^-}$.
\par
If one would simply eliminate all events with
secondary radiation of hadrons and leptons, one would arrive at
a measurement of $\Gamma(Z \to \tau^+ \tau^-)$ significantly
below (about 0.4\%) the generally quoted inclusive value.
Both real and virtual radiation are enhanced by the square of the large
logarithm $\ln(M_Z^2/m_{had}^2)$ times $\ln(M_Z^2/m_l^2)\,$.
Because the coefficients of these logarithms only differ by sign,
real and virtual corrections cancel to a large extent. This leads to a
small contribution to the inclusive decay rate, as can be seen in the next
section. Measurements of exclusive $\tau$ pair production, however, are
strongly affected. With this application in mind we proceed to the
calculation of the virtual \pagebreak[4] corrections.
\par
\vspace{1cm}\noindent
{\bf 3.~Virtual radiation of hadrons}\\
\par
\noindent
Virtual corrections to $\tau$ pair production affect both Dirac and
Pauli form factors, $F_1$ and $F_2$. Leptonic corrections have been
calculated in analytic form in \cite{HKT1} for the cases that the
masses of the radiated leptons are much smaller than or equal to
the $\tau$ mass. The effect of virtual radiation of hadrons via
hadronic vacuum polarisation of one exchanged photon, as depicted in
Fig.\ref{figgraphs}b, is given
by an expression similar to eq.~(\ref{masterreal}).
For virtual radiation the phase space integration has to be
extended to infinity corresponding to the dispersion integration of
the absorptive part of the hadronic vacuum polarisation.
The function $F(s',s)$, now representing the normalized cross
section for the production of a $\tau$ pair with an exchanged vector
boson of mass $\sqrt{s'}$, may be easily calculated analytically
and reads
\begin{equation}
F(s',s) =  \left(\frac{\alpha}{\pi}\right)^2\,
\bigg[\,\omega(3-\omega^2)\left[
{\rm Re}\widehat{F}_1(s',s)+{\rm Re}\widehat{F}_2(s',s)
\right] +
\omega^3\,{\rm Re}\widehat{F}_2(s',s)\,\bigg]\,,
\label{fhatvirt}
\end{equation}
where
\begin{equation}
\omega = \sqrt{1-4m_\tau^2/s}\,.
\label{omega}
\end{equation}
The functions ${\rm Re}\widehat{F}_{1,2}$ are
given in~\cite{HKT1} (eqs. (15) and (16)) and denote the real parts of the
electromagnetic form factors arising from the exchange of a vector
boson of mass $\sqrt{s'}$. The integral
\begin{equation}
R_{\tau^+\tau^-\,had}^{virt} =
 \frac{1}{3} \,
 \left(\frac{\alpha}{\pi}\right)^2\,
 \int\limits_{4m_{\pi}^2}^{\infty} \frac{{\rm d}s'}{s'}
 \, R_{had}(s') \, F(s',s)
\label{mastervirt}
\end{equation}
is calculated numerically. A similar formula applies to
the form factors individually.
\par
\begin{figure}
\begin{center}
\leavevmode
\epsfxsize=12cm
\epsffile[120 300 450 530]{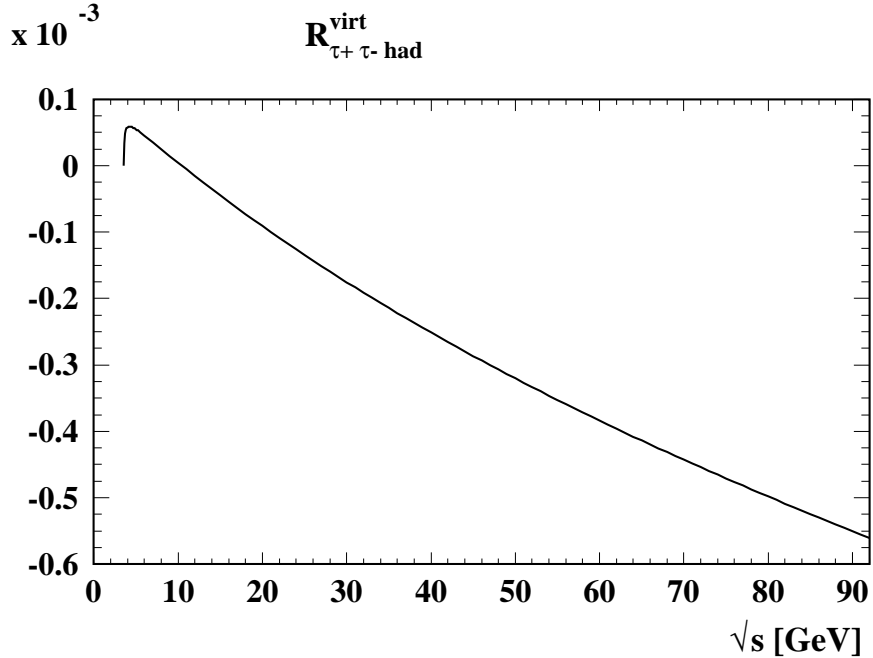}
\vskip -3mm
\caption[]{{\em Contribution to $e^+ e^- \to \tau^+ \tau^-$ due to virtual
radiation of hadrons.}}
\label{scanv}
\end{center}
\end{figure}

Virtual corrections from different hadronic intermediate states cannot
be separated experimentally and contribute in a coherent manner. Their
combined effect on the form factors and the rate are displayed in
Table~\ref{tab2}
for four characteristic energies.
\begin{table}
\begin{center}
\begin{tabular}{|c||c|c|c|c|} \hline
$\sqrt{s}$
& $3.4552$ GeV $= 2\,m_{\tau}$ & $4.2$ GeV & $10$ GeV & $M_Z$ \\ \hline \hline
$F_{1, had}$ & $7.45\cdot 10^{-5}$ & $3.55\cdot 10^{-5}$ &
               $2.69\cdot 10^{-6}$ & $-2.78\cdot 10^{-4}$ \\ \hline
$F_{2, had}$ & $9.00\cdot 10^{-6}$ & $4.68\cdot 10^{-6}$ &
               $-3.21\cdot 10^{-7}$ & $-6.03\cdot 10^{-8}$ \\ \hline
$R_{\tau^+ \tau^- had}^{virt}$
             & $0$ & $5.89\cdot 10^{-5}$ &
               $4.45\cdot 10^{-6}$ & $-5.56\cdot 10^{-4}$ \\ \hline \hline
$F_{1, e^+ e^-}$ & $1.50\cdot 10^{-2}$ & $5.84\cdot 10^{-5}$ &
                   $-1.82\cdot 10^{-4}$ & $-8.94\cdot 10^{-4}$ \\ \hline
$F_{2, e^+ e^-}$ & $2.30\cdot 10^{-3}$ & $-4.26\cdot 10^{-6}$ &
                   $-2.43\cdot 10^{-6}$ & $-8.21\cdot 10^{-8}$ \\ \hline
$R_{\tau^+ \tau^- e^+ e^-}^{virt}$
                 & $0$ & $7.78\cdot 10^{-5}$ &
                   $-3.68\cdot 10^{-4}$ & $-1.79\cdot 10^{-3}$ \\
\hline \hline
$F_{1, \mu^+ \mu^-}$ & $5.86\cdot 10^{-5}$ & $1.76\cdot 10^{-5}$ &
                       $-5.98\cdot 10^{-6}$ & $-1.38\cdot 10^{-4}$ \\ \hline
$F_{2, \mu^+ \mu^-}$ & $7.74\cdot 10^{-6}$ & $1.74\cdot 10^{-6}$ &
                       $-3.37\cdot 10^{-7}$ & $-2.57\cdot 10^{-8}$ \\ \hline
$R_{\tau^+ \tau^- \mu^+ \mu^-}^{virt}$
                     & $0$ & $2.82\cdot 10^{-5}$ &
                       $-1.28\cdot 10^{-5}$ & $-2.77\cdot 10^{-4}$ \\
\hline \hline
$F_{1, \tau^+ \tau^-}$ & $1.25\cdot 10^{-6}$ & $1.31\cdot 10^{-6}$ &
                         $2.16\cdot 10^{-6}$ & $-1.08\cdot 10^{-5}$ \\ \hline
$F_{2, \tau^+ \tau^-}$ & $1.03\cdot 10^{-7}$ & $1.14\cdot 10^{-7}$ &
                         $6.00\cdot 10^{-8}$ & $-3.73\cdot 10^{-9}$ \\ \hline
$R_{\tau^+ \tau^- \tau^+ \tau^-}^{virt}$
                       & $0$ & $2.08\cdot 10^{-6}$ &
                         $4.47\cdot 10^{-6}$ & $-2.16\cdot 10^{-5}$ \\
\hline
\end{tabular}
\caption[]{\label{tab2}
{\em Virtual corrections due to hadronic and leptonic vacuum
polarisation for center of mass energies
$\sqrt{s} = 2\,m_{\tau}/4.2$ GeV/$10$ GeV/$M_Z$ for form factors and
the resulting cross section as described in the text.}}
\end{center}
\end{table}
For completeness we also display the contributions from leptonic vacuum
polarisation. Contributions from virtual electron pairs are obtained
with analytical formulae derived in the limit $m_e \ll m_{\tau}$ (see
\cite{HKT1}).
In the same reference an analytic formula for virtual $\tau$ pairs
is also at hand. It is evident from Table~\ref{tab2}
that these corrections must be taken into
consideration in precision measurements of partial rates.
Close to threshold the virtual corrections give a positive
contribution to $R$ as is illustrated in Fig.~\ref{scanv}, where
$R_{\tau^+ \tau^- had}^{virt}$ is plotted for the range
$2\,m_{\tau} < \sqrt{s} < M_Z$. This is a consequence of the
Coulomb attraction mediated by the exchange of a vector boson.
Far above threshold the corrections are negative. They lead to
a reduction of the exclusive decay rate of $Z \to \tau^+ \tau^-$ by
nearly three per mill, exceeding the ${\cal O}(\alpha)$ correction
of $3/4(\alpha/\pi) \approx 1.7 \cdot 10^{-3}$. As mentioned above
these negative corrections cancel for the most part
the positive corrections from real emission, leaving a small positive
overall correction. This remainder can to a large extent be
absorbed into a running
electromagnetic coupling constant $\alpha(s)$ as discussed in~\cite{HKT1}.
\par
\vspace{1cm}\noindent
{\bf 4.~Corrections to $Z$ decays into $\mu^+ \mu^-$ and $e^+ e^-$}\\
\par
\noindent
In \cite{KKKS,HJKT1} analytical and numerical results were presented for
real and virtual radiation of massive leptons or hadrons from the primary
production of massless leptons. For completeness it seems appropriate to
present in this paper a brief update of the hadronic radiation in $Z$
decays to $\mu^+ \mu^-$ and $e^+ e^-$ together with a complete table
for the radiation of leptons.
Real and virtual radiation are presented in Tables~\ref{tab3} and
\ref{tab4} respectively.
It is evident that
exclusive determinations of $Z \to e^+ e^-$ or $\mu^+ \mu^-$ will
become sensitive towards these corrections.
\par
The results for $R_{e^+ e^- had}$ and
$R_{\mu^+ \mu^- had}$ are $\approx$6\% larger than our old
value \cite{HJKT1}. Half of this
effect is due to the updated description of the continuum and 3\%
originate from the inclusion of the effective electromagnetic coupling
constant $\alpha$ (see also the appendix).
Comparing our result for $R_{e^+ e^- had}^{virt}$ (taking
$\sqrt{s} = 93$ GeV) with the corresponding one in \cite{KKKS} the
difference amounts to only 1\%.
Our results for the radiation of leptons are about 15\% lower than
those of~\cite{vdB}, essentially a consequence of the different
choice for $\alpha$. There the running electric coupling constant
$\alpha(M_Z^2)$ was chosen. However, radiation of leptons, in
particular of electrons, is dominated by small $s^\prime$. Therefore
the low energy scale adopted in this work is more appropriate.
\begin{table}
\begin{center}
\begin{tabular}{|c||c|c|c|} \hline
secondary$\backslash$primary
        & $e$                 & $\mu$               & $\tau$
\\ \hline \hline
$e$     & $2.78\cdot 10^{-3}$ & $2.53\cdot 10^{-3}$ & $1.82\cdot 10^{-3}$
\\ \hline
$\mu$   & $3.32\cdot 10^{-4}$ & $3.31\cdot 10^{-4}$ & $2.94\cdot 10^{-4}$
\\ \hline
$\tau$  & $3.23\cdot 10^{-5}$ & $3.21\cdot 10^{-5}$ & $3.13\cdot 10^{-5}$
\\ \hline \hline
$hadrons$
        & $6.68\cdot 10^{-4}$ & $6.67\cdot 10^{-4}$ & $6.14\cdot 10^{-4}$
\\ \hline \hline \hline
   Sum     & $3.81\cdot 10^{-3}$ & $3.56\cdot 10^{-3}$ & $2.76\cdot 10^{-3}$
\\ \hline
\end{tabular}
\caption[]{\label{tab3}
{\em Real radiation $R_{primary, secondary}$ for $s = M_Z^2$.}}
\end{center}
\end{table}
\begin{table}
\begin{center}
\begin{tabular}{|c||c|c|c|} \hline
secondary$\backslash$primary
        & $e$                  & $\mu$                 & $\tau$
\\ \hline \hline
$e$     & $-2.75\cdot 10^{-3}$ & $-2.50\cdot 10^{-3}$ & $-1.79\cdot 10^{-3}$
\\ \hline
$\mu$   & $-3.15\cdot 10^{-4}$ & $-3.14\cdot 10^{-4}$ & $-2.77\cdot 10^{-4}$
\\ \hline
$\tau$  & $-2.24\cdot 10^{-5}$ & $-2.24\cdot 10^{-5}$ & $-2.16\cdot 10^{-5}$
\\ \hline \hline
$hadrons$
        & $-6.09\cdot 10^{-4}$ & $-6.09\cdot 10^{-4}$ & $-5.56\cdot 10^{-4}$
\\ \hline \hline \hline
   Sum  & $-3.69\cdot 10^{-3}$ & $-3.44\cdot 10^{-3}$ & $-2.64\cdot 10^{-3}$
\\ \hline
\end{tabular}
\caption[]{\label{tab4}
{\em Virtual radiation $R_{primary, secondary}^{virt}$ for $s = M_Z^2$.}}
\end{center}
\end{table}
\par
Below we also list the high energy expansions (neglecting terms of
order $m_{1/2}^2/s$) of $R_{f_1 \bar f_1 f_2 \bar f_2}$ and
$R_{f_1 \bar f_1 f_2 \bar f_2}^{virt}$ for the cases
$m_1 \gg m_2$, $m_1 = m_2$ and $m_1 \ll m_2$, where $m_1$ is
the mass of the primary produced fermion $f_1$ and $m_2$ the mass of the
secondary emitted fermion $f_2$. These formulae provide excellent
approximations for $\sqrt{s} = M_Z$ (better than one per mill in all
cases except for $f_2 = \tau$, where the relative difference between
complete and approximate result amounts to nearly $1\%$) and
therefore can directly be applied to the corresponding leptonic $Z$
partial widths.
\begin{eqnarray}
R_{f_1\bar f_1 f_2\bar f_2}^{\ m_1 \gg m_2} & = &
 \left(\frac{\alpha}{\pi}\right)^2\,
 \bigg[ - \,\ln^2 \frac{m_2^2}{s}\,\frac{1}{6}\,
     \bigg( \ln \frac{m_1^2}{s} + 1 \bigg) \,\nonumber\,
     \\
 & & \mbox{}\,\qquad\quad    +
  \ln \frac{m_2^2}{s}\,\bigg( \frac{1}{6}\,
      \ln^2 \frac{m_1^2}{s} -
     \frac{13}{18}\,\ln \frac{m_1^2}{s} +
     \frac{4}{3}\,\zeta(2) - \frac{53}{36} \bigg) \,
   \nonumber\,\\
 & & \mbox{}\,\qquad\quad  -
  \frac{1}{18}\,\ln^3 \frac{m_1^2}{s} +
  \frac{13}{36}\,\ln^2 \frac{m_1^2}{s} -
  \left( \frac{133}{108} + \frac{2}{3}\,\zeta(2) \right) \,
   \ln \frac{m_1^2}{s}\,\nonumber\,\\
 & & \mbox{}\,\qquad\quad
   + \frac{5}{3}\,\zeta(3) + \frac{32}{9}\,\zeta(2) -
  \frac{833}{216} - \frac{3}{4}\,\frac{m_2}{m_1}\,{{\pi }^2} \,\bigg]\,
\label{mismallreal} \\
R_{f_1\bar f_1 f_2\bar f_2}^{\ m_1 = m_2} & = &
 \left(\frac{\alpha}{\pi}\right)^2\,
 \bigg[ - \frac{1}{18}\,\ln^3 \frac{m_2^2}{s}  -
  \frac{19}{36}\,\ln^2 \frac{m_2^2}{s} +
  \frac{2}{3}\,\left( -\frac{73}{18} + \zeta(2) \right) \,
   \ln \frac{m_2^2}{s}\,\nonumber\,\\
 & & \mbox{}\,\qquad\quad
   + \zeta(3) + \frac{11}{3}\,\zeta(2) - \frac{1829}{216} \,\bigg]\,
\label{miequalreal} \\
R_{f_1\bar f_1 f_2\bar f_2}^{\ m_1\ll m_2} & = &
 \left(\frac{\alpha}{\pi}\right)^2\,
 \bigg[ - \frac{1}{18}\,\ln^3 \frac{m_2^2}{s}  -
  \frac{19}{36}\,\ln^2 \frac{m_2^2}{s} +
  \frac{2}{3}\,\left( -\frac{73}{18} + \zeta(2) \right) \,
   \ln \frac{m_2^2}{s}\,\nonumber\,\\
 & & \mbox{}\,\qquad\quad
   + \frac{5}{3}\,\zeta(3) + \frac{19}{9}\,\zeta(2) -
  \frac{2123}{324} \,\bigg]\,
\label{milargereal}
\end{eqnarray}
\pagebreak
\begin{eqnarray}
R_{f_1\bar f_1 f_2\bar f_2}^{\ m_1 \gg m_2, virt} & = &
 \left(\frac{\alpha}{\pi}\right)^2 \,
 \bigg[ \ln^2 \frac{m_2^2}{s}\,\frac{1}{6}\,
   \bigg( \ln \frac{m_1^2}{s} + 1 \bigg) \,\nonumber\,
   \\
 & & \mbox{}\,\qquad\quad  + \ln \frac{m_2^2}{s}\,
   \bigg( - \frac{1}{6}\,\ln^2 \frac{m_1^2}{s}
       + \frac{13}{18}\,\ln \frac{m_1^2}{s} -
     \frac{4}{3}\,\zeta(2) + \frac{11}{9} \bigg) \,
   \nonumber\,\\
 & & \mbox{}\,\qquad\quad  +
  \frac{1}{18}\,\ln^3 \frac{m_1^2}{s} -
  \frac{13}{36}\,\ln^2 \frac{m_1^2}{s} +
  \left( \frac{133}{108} + \frac{2}{3}\,\zeta(2) \right) \,
   \ln \frac{m_1^2}{s}\,\nonumber\,\\
 & & \mbox{}\,\qquad\quad
   - \frac{2}{3}\,\zeta(3) - \frac{32}{9}\,\zeta(2) +
  \frac{67}{27} + \frac{3}{4}\,\frac{m_2}{m_1}\,{{\pi }^2} \,\bigg]\,
\label{mismallvirt} \\
R_{f_1\bar f_1 f_2\bar f_2}^{\ m_1 = m_2, virt} & = &
 \left(\frac{\alpha}{\pi}\right)^2\,
 \bigg[ \frac{1}{18}\,\ln^3 \frac{m_2^2}{s} +
  \frac{19}{36}\,\ln^2 \frac{m_2^2}{s} +
  \frac{2}{3}\,\left( \frac{265}{72} - \zeta(2) \right) \,
   \ln \frac{m_2^2}{s} - \frac{11}{3}\,\zeta(2) +
  \frac{383}{54} \,\bigg]\,
\label{miequalvirt} \\
R_{f_1\bar f_1 f_2\bar f_2}^{\ m_1\ll m_2, virt} & = &
 \left(\frac{\alpha}{\pi}\right)^2\,
 \bigg[ \frac{1}{18}\,\ln^3 \frac{m_2^2}{s} +
  \frac{19}{36}\,\ln^2 \frac{m_2^2}{s} +
  \frac{2}{3}\,\left( \frac{265}{72} - \zeta(2) \right) \,
   \ln \frac{m_2^2}{s}\,\nonumber\,\\
 & & \mbox{}\,\qquad\quad
   - \frac{2}{3}\,\zeta(3) - \frac{19}{9}\,\zeta(2) +
  \frac{3355}{648} \,\bigg]\,
\label{milargevirt}
\end{eqnarray}
Results for virtual corrections were first obtained
in~\cite{HKT1}, \cite{BURGERS} and \cite{KKKS} for the
cases $m_1\gg m_2, m_1=m_2$ and
$m_1\ll m_2$, respectively. Those for real radiation are taken
from~\cite{HKT1} for $m_1\gg m_2$ and from~\cite{HJKT1,HJKT2} for
$m_1\ll m_2$. The formula for real radiation with $m_1=m_2$ is new.
The leading cubic logarithms for all these cases can also be found
in~\cite{BFK}.
Note that in (\ref{mismallreal}) and (\ref{mismallvirt}) linear mass
corrections $\sim m_2/m_1$ are included. They significantly improve the
quality of the approximations, especially in the case
$m_1 = m_{\tau}$, $m_2 = m_{\mu}$. Neglecting these linear mass corrections
the value of say $R_{\tau^+ \tau^- \mu^+ \mu^-}^{\ m_1 \gg m_2, virt}$
for the energy $\sqrt{s} = 20$ GeV differs from the value based on
numerical integration by 4\%, while including them leads to a
difference of only three per mill. Of course, for higher energies
the latter difference decreases and amounts to eight per mill
vs. half a per mill for $\sqrt{s} = M_Z$. There are no linear mass
corrections to the other high energy approximations.
It is an interesting fact that the complete results for these linear
mass corrections, valid for all energies above threshold, are
remarkably simple:
\begin{eqnarray}
R_{f_1 \bar f_1 f_2 \bar f_2, linear}^{m_1 \gg m_2} &=&
-\frac{3}{8}\,\omega\,(3-\omega^2)\,\frac{m_2}{m_1}\,\pi^2\,,
\label{linearreal}\\
R_{f_1 \bar f_1 f_2 \bar f_2, linear}^{m_1 \gg m_2, virt} &=&
\frac{3}{8}\,\frac{1+\omega^2}{\omega}\,\frac{m_2}{m_1}\,\pi^2
\label{linearvirt}
\end{eqnarray}
with $\omega$ defined as in (\ref{omega}).
They are in very close relation to the non--analytic linear vector boson
mass terms of the corresponding one loop results denoted by $F$,
eqs.~(\ref{masterreal}) and (\ref{fhatvirt}). The difference between
(\ref{linearreal}) and (\ref{linearvirt}) can be reproduced by the
analogous linear mass correction in the two loop relation between
pole and running masses, see \cite{Gray}.
\par
Again, concerning the total $Z$ decay rate, it is obvious from
eq.~(\ref{mismallreal})--(\ref{milargevirt}) that by adding the
corresponding real and virtual contributions the dominant
logarithmic terms cancel, leaving the result
\begin{equation}
R_{f_1 \bar f_1 f_2 \bar f_2} + R_{f_1 \bar f_1 f_2 \bar f_2}^{virt} =
 \left(\frac{\alpha}{\pi}\right)^2\,
 \left[\,\frac{1}{4}\,\ln\frac{s}{m_2^2} + \zeta(3) - \frac{11}{8}\,\right]
\label{rsumrealvirthigh}
\end{equation}
which is universal for all three mass constellations in the high energy
limit. As mentioned in the previous section the remaining logarithm can be
absorbed into the running coupling constant $\alpha(s)$, leaving a tiny term
$\sim (\alpha/\pi)^2$ without logarithm.
\par
For completeness we also present the prediction for the radiation of
hadrons from primary electrons and muons. In this case the result can be
expressed in terms of the moments
\begin{equation}
R_n = \int_0^1 \frac{{\rm d}x}{x} \frac{\ln^nx}{n!}
\left(R_{had}(4m^2/x)-R_{had}(\infty)\right)
\label{moments}
\end{equation}
leading to the following formulae for real~\cite{HJKT1} and for
virtual~\cite{KKKS} radiation
($l=e,\mu$)
\pagebreak
\begin{eqnarray}
\lefteqn{ R_{l^+l^-had}  = }\nonumber\\[1mm]
& & \frac{1}{6} \left(\frac{\alpha}{\pi}\right)^2 \biggl\{
R_{had}(\infty) \left[\, \frac{1}{3} \ln^3\frac{s}{4m_\pi^2} -
\frac{3}{2} \ln^2\frac{s}{4m_\pi^2} +
\left(-2\,\zeta(2)+5\right)\ln \frac{s}{4m_\pi^2}+
\left(6\,\zeta(3)+3\,\zeta(2)-\frac{19}{2}\right)
\,\right]\nonumber\\[1mm]
& & \qquad\qquad+\,R_0 \left[\,\ln^2\frac{s}{4m_\pi^2}-
3\ln \frac{s}{4m_\pi^2}-2\,\zeta(2)+5\,\right] +
R_1 \left[\,2\ln \frac{s}{4m_\pi^2}-3\,\right] + 2\, R_2 \biggr\}\,,
\label{momreal} \\
\lefteqn{ R_{l^+l^-had}^{virt}  = }\nonumber\\[1mm]
& & \frac{1}{6} \left(\frac{\alpha}{\pi}\right)^2 \biggl\{
R_{had}(\infty) \left[\, -\frac{1}{3}\ln^3\frac{s}{4m_\pi^2}+\frac{3}{2}
\ln^2\frac{s}{4m_\pi^2}+\left(2\,\zeta(2)-
\frac{7}{2}\right)\ln \frac{s}{4m_\pi^2}-3
\,\zeta(2)+\frac{15}{4}\,\right] \nonumber\\[1mm]
& & \qquad\qquad+\,R_0\left[\,-\ln^2\frac{s}{4m_\pi^2}+
3\ln \frac{s}{4m_\pi^2}+ 2\,\zeta(2)-
\frac{7}{2}\,\right]+R_1 \left[\,-2\ln \frac{s}{4m_\pi^2}+
3\,\right] - 2\,R_2 \biggr\}\,.
\label{momvirt}
\end{eqnarray}
The evaluation of (\ref{moments}) gives
\begin{eqnarray}
R_{had}(\infty) &=&
 \;\;\;\;\, 4.31\,,\nonumber\\
R_0 &=& -11.15\,,\nonumber\\
R_1 &=& \;\;\: 25.05\,,\nonumber\\
R_2 &=& -48.00
\label{momnum}
\end{eqnarray}
where we used the dressed $R_{had}(s')$ including the narrow resonances
and $\sqrt{s'} = 100$ GeV as the integration cutoff. The values of the
moments individually depend on the choice of this cutoff (and
correspondingly $R_{had}(\infty)$), but we have checked that the
results of (\ref{momreal}) and
(\ref{momvirt}) are identical for all cutoff energies above $40$ GeV.
\begin{figure}
\begin{center}
\leavevmode
\epsfxsize=8.6cm
\epsffile[175 335 400 515]{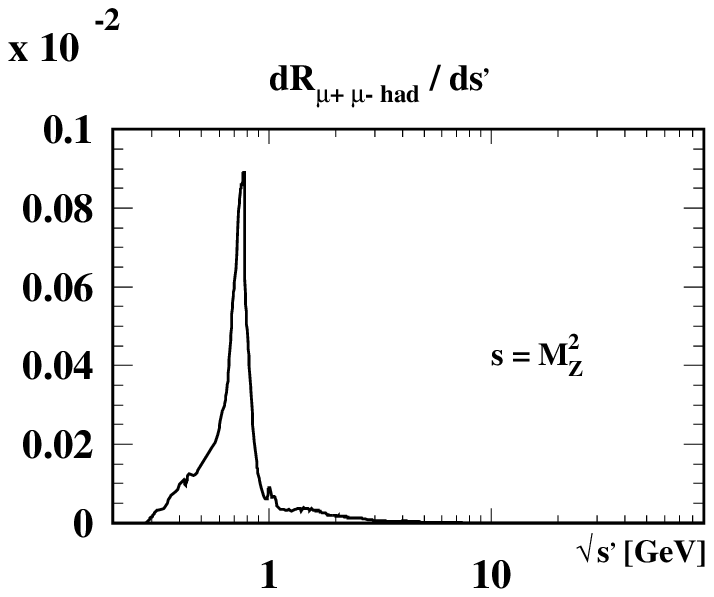}
\leavevmode
\epsfxsize=8.6cm
\epsffile[175 335 400 515]{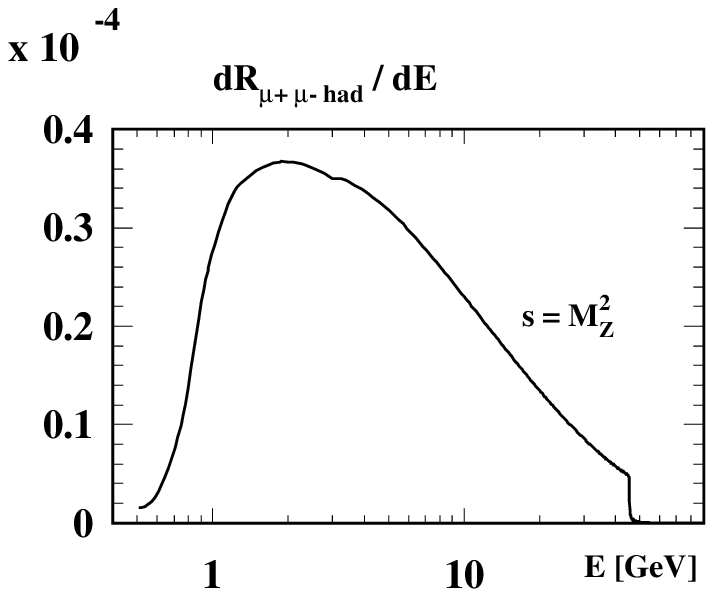}\\
\leavevmode
\epsfxsize=8.6cm
\epsffile[175 335 400 515]{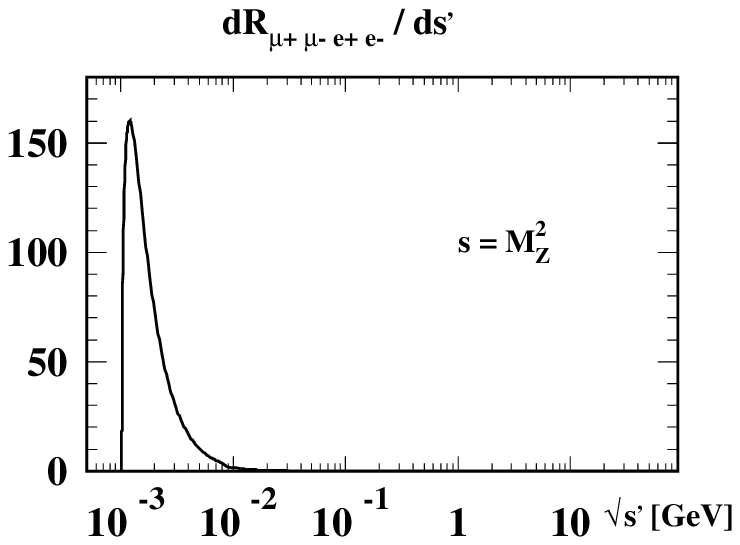}
\leavevmode
\epsfxsize=8.6cm
\epsffile[175 335 400 515]{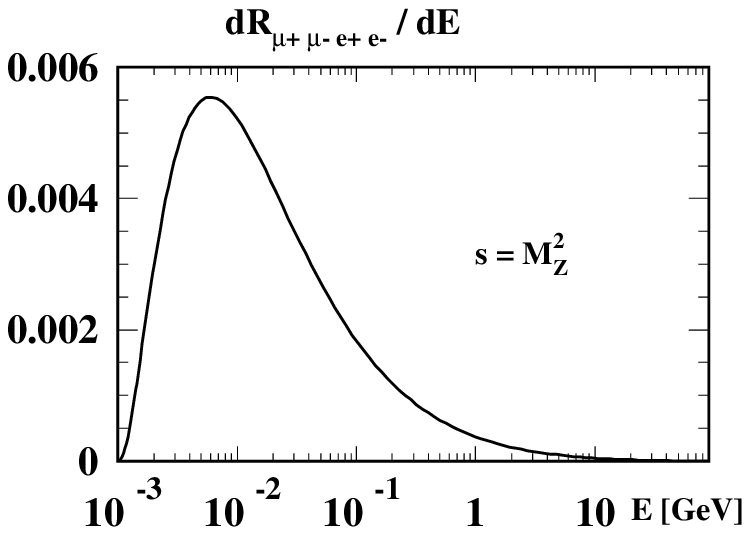}
\vskip -3mm
\caption[]{{\em Differential distributions for real emission of
hadrons and of $e^+e^-$ pairs in $\mu$ pair production for $\sqrt{s} = M_Z$.}}
\label{figmuhadedif}
\end{center}
\end{figure}
\noindent
For the derivation of eqs.~(\ref{momreal}) and (\ref{momvirt}) the
approximation $m_{l}\ll m_\pi$ was essential. The relative difference
between the asymptotic formulae and the complete numerical results is at
most 0.2\% for $\sqrt{s} = M_Z$. For lower energies, of course, the
difference increases. Similar formulae for
radiation off the primary tau pair provide a poor
approximation to the true answer for real and
virtual radiation individually, but are adequate for the sum, which
is identical to the electron and muon case (see also Tables~\ref{tab3} and
\ref{tab4}).
For the sum of real and virtual radiation of leptons and hadrons at
$\sqrt{s} = M_Z$, which is valid for all three lepton channels
$f_1 = e, \mu, \tau$, one arrives at
\begin{eqnarray}
\lefteqn{R_{f_1 \bar f_1 had} + R_{f_1 \bar f_1 had}^{virt} +
\sum_{f_2=e, \mu, \tau} \left(\,R_{f_1 \bar f_1 f_2 \bar f_2} +
                   R_{f_1 \bar f_1 f_2 \bar f_2}^{virt}\,\right)}
\nonumber\\
&=&
 \frac{3}{4}\,\bigg[\,\frac{\alpha(M_Z^2)}{\pi} - \frac{\alpha}{\pi}\,\bigg]
 + \left(\frac{\alpha}{\pi}\right)^2\,
   \left[\,\zeta(3) - \frac{23}{24}\,\right]\,
 \bigg(\,R_{had}({\infty})+\sum_{f_2=e, \mu, \tau}Q^2_{f_2}\,\bigg)
\nonumber \\
&\approx& 0.12\cdot 10^{-3}\,,
\label{sumrealplusvirt}
\end{eqnarray}
where the usual definition of the running electromagnetic
coupling constant
\begin{eqnarray}
\frac{1}{\alpha(s)} - \frac{1}{\alpha} = \Pi(s)
\label{runalphadef}
\end{eqnarray}
was employed. Here the hadronic contributions to the photon vacuum
polarisation have also been parametrized by the moments defined
in (\ref{moments}):
\begin{eqnarray}
\Pi(s)
 \stackrel{s \to \infty}{\longrightarrow}
 - \frac{1}{3\,\pi}\,
 \bigg[\,\sum_{i = e, \mu, \tau}\,\bigg(\,\ln\frac{s}{m_i^2} -
 \frac{5}{3}\,\bigg) + R_{had}(\infty) \ln\frac{s}{4\,m_{\pi}^2} + R_0
\,\bigg]\,.
\label{runalpha}
\end{eqnarray}
\par
The differential distribution
${\rm d} R_{l^+l^- had}/{\rm d} s_{\scriptsize had}^\prime$
for the case of massless leptons is given in closed form
(see eqs.~(1) and (2) of~\cite{HJKT1}),
${\rm d} R_{l^+l^- had}/{\rm d} E_{\scriptsize had}$,
of course, has to be calculated numerically. The predicted distributions
for the radiation of hadrons and of $e^+e^-$ pairs off primary produced muon
pairs are shown\footnote{The figure for
${\rm d} R_{l^+l^- had}/{\rm d} s_{\scriptsize had}^\prime$ does not
include the narrow resonances. The missing contributions
to $R_{l^+l^- had}$
are $\omega$: $2.33\cdot 10^{-5}$, $\phi$: $3.56\cdot 10^{-5}$,
$J/\psi$: $2.95\cdot 10^{-5}$ and $\Upsilon$: $8.33\cdot 10^{-7}\,$.}
in Fig.~\ref{figmuhadedif}. Again
the dominant configuration
consists of an $\mu^+\mu^-$ with invariant mass fairly close
to $M_Z$ and a soft hadronic or $e^+e^-$ system.
\par
\vspace{1cm}\noindent
{\bf 5.~Summary}\\
\par
\noindent
A comprehensive treatment of the radiation of hadronic final
states from a primary $\tau$ pair is presented, which is
applicable over the full energy range of interest  in the
foreseeable future. In the low energy region real as well as
virtual radiation is small, typically at the level of
$10^{-4}$ compared to the primary reaction. The situation is
entirely different for a c.m. energy around $90$ GeV. In this
case important differences arise between ``inclusive''  and
``exclusive'' measurements of the $Z\to \tau^+\tau^-$ rate,
in particular if radiation of the various hadronic and
fermionic channels is collected, or eliminated by experimental cuts.
We have presented differential distributions with respect to
the invariant mass and the energy of the hadronic system, and for
comparison the same quantities for the radiation of
electron--positron pairs. It is evident that all
distributions are peaked at small masses and energies
respectively. The calculation of the rate for hadronic
radiation must necessarily rely on a numerical integration
of experimental data. Leptonic radiation, on the other hand,
can be calculated analytically for the mass assignments of
interest. Handy formulae are presented for different
mass assignments which are applicable in the high energy
limit, and which are valid up to terms of order
$m_{lept}^2/s$. They may be useful also for other
applications.
\par
\vspace{1cm}\noindent
{\em Acknowledgment:}
We would like to thank F.~Jegerlehner and S.~Eidelman for providing
their compilation of data and discussing its proper treatment. We are
also grateful to K.~Chetyrkin, W.~Hollik and D.~Schaile for helpful
discussions. One of the authors (J.H.K.) would like to thank the SLAC
theory group for hospitality, and the Volkswagen--Stiftung
grant I/70 453 for generous support.
\par
\vspace{1cm}\noindent
{\bf Appendix: $R_{had}(s')$}\\
\par
\noindent
In our calculation we used experimental data for $R_{had}(s')$ as
presented in a recent publication by F.~Jegerlehner and
S.~Eidelman\cite{jegerl}. For a detailed discussion we refer the reader
to this work.
\par
In the region close to the $\pi^+ \pi^-$ threshold where no experimental
data are available we use the chiral expansion of the pion form factor
(eq.~(18) of \cite{jegerl}).
The continuum-- and $\rho$--contributions are parametrized by weighted
averages of experimental data from various experiments. From $318$~MeV up
to $810$~MeV these are given in terms of the pion form
factor $|F_{\pi}|^2$. The continuum from $810$~MeV up to $40$~GeV
(above which we use the prediction of perturbation theory) is well
described by weighted averages of $R_{had}(s)$. At this point two comments
are in order: The higher resonances $\psi(4040),\ \psi(4160)$\ and
$\psi(4415)$ are included in the weighted averages of the continuum data
(as can be seen in Fig.~7 of \cite{jegerl}) whereas the resonances
$\omega$, $\Phi$, $J/\psi(1S)$, $\psi(2S)$, $\psi(3770)$ and the
$\Upsilon$--family (six resonances) are treated in the narrow width
approximation as described above.\\
\par
All experimental values of $R_{had}$ (and $|F_{\pi}|^2$) are given
as ``undressed'' ($\equiv$ lowest order in QED) quantities. In
order to resum all additional self--energy insertions in the photon
propagator we
use for the integration of (\ref{masterreal}), (\ref{diffenergyreal})
and (\ref{mastervirt}) the ``dressed'' quantity
\begin{equation}
R_{had}^{dressed}(s) =
\left(\frac{\alpha(s)}{\alpha}\right)^2
R_{had}(s)\,,\quad
\alpha(s)=\frac{\alpha}{1-\Delta\alpha}\,.
\label{rdressed}
\end{equation}
In the low energy region up to $\sqrt{s'} = 3$\ GeV most experiments
only used the leptonic contribution of $\Delta\alpha$ to extract
one--particle--irreducible quantities from the measured cross--sections.
For that reason we take $\Delta\alpha = \Delta\alpha_{lep}$ in
this region. For all other energies
$$\Delta\alpha = \Delta\alpha_{lep} + \Delta\alpha_{had}\,.$$
As mentioned above these corrections led to an increase of the
result by a factor of $\approx 1.03$.
We also calculated the effect of $\Delta\alpha_{lep}(s)$ on our results,
comparing the full one loop expression of the fermionic part of the
photon self energy with the leading logarithmic behaviour (as it appears
in the renormalization group ``running $\alpha$''). The difference of
$\approx$0.5\% may be considered as a crude estimate of other effects
not taken into account via the implicit resummation by
using $\alpha(s)$ and is smaller than
the claimed accuracy of our calculation.\\
\vskip 1.5cm
\begin{sloppy}
\begin{raggedright}
\def\app#1#2#3{{\it Act. Phys. Pol. }{\bf B #1} (#2) #3}
\def\apa#1#2#3{{\it Act. Phys. Austr.}{\bf #1} (#2) #3}
\def\lhc{Proc. LHC Workshop, CERN 90-10}
\def\npb#1#2#3{{\it Nucl. Phys. }{\bf B #1} (#2) #3}
\def\plb#1#2#3{{\it Phys. Lett. }{\bf B #1} (#2) #3}
\def\prd#1#2#3{{\it Phys. Rev. }{\bf D #1} (#2) #3}
\def\pR#1#2#3{{\it Phys. Rev. }{\bf #1} (#2) #3}
\def\prl#1#2#3{{\it Phys. Rev. Lett. }{\bf #1} (#2) #3}
\def\prc#1#2#3{{\it Phys. Reports }{\bf #1} (#2) #3}
\def\cpc#1#2#3{{\it Comp. Phys. Commun. }{\bf #1} (#2) #3}
\def\nim#1#2#3{{\it Nucl. Inst. Meth. }{\bf #1} (#2) #3}
\def\pr#1#2#3{{\it Phys. Reports }{\bf #1} (#2) #3}
\def\sovnp#1#2#3{{\it Sov. J. Nucl. Phys. }{\bf #1} (#2) #3}
\def\jl#1#2#3{{\it JETP Lett. }{\bf #1} (#2) #3}
\def\jet#1#2#3{{\it JETP Lett. }{\bf #1} (#2) #3}
\def\zpc#1#2#3{{\it Z. Phys. }{\bf C #1} (#2) #3}
\def\ptp#1#2#3{{\it Prog.~Theor.~Phys.~}{\bf #1} (#2) #3}
\def\nca#1#2#3{{\it Nouvo~Cim.~}{\bf #1A} (#2) #3}
{\elevenbf\noindent  References \hfil}
\vglue 0.4cm

\end{raggedright}
\end{sloppy}
\end{document}